\documentclass[12pt]{iopart}
\usepackage{accents} 
\usepackage{amssymb}

\def\thebibliography#1{\list
 {\hfil[\arabic{enumi}]}{
 \settowidth\labelwidth{\footnotesize[#1]}%
 \leftmargin\labelwidth
 \advance\leftmargin\labelsep
 \advance\leftmargin -\itemindent
 \usecounter{enumi}}\footnotesize
 \def\newblock{\ }
 \sloppy\clubpenalty4000\widowpenalty4000
 \sfcode`\.=1000\relax
}
\def\m@th{\mathsurround=0pt}
\mathchardef\bracell="0365 
\def\upbrall{$\m@th\bracell$}
\def\undertilde#1{\mathop{\vtop{\ialign{##\crcr
    $\hfil\displaystyle{#1}\hfil$\crcr
     \noalign
     {\kern1.5pt\nointerlineskip}
     \upbrall\crcr\noalign{\kern-7pt
   }}}}\limits}
\def\underdot#1{\mathop{\vtop{\ialign{##\crcr
    $\hfil\displaystyle{#1}\hfil$\crcr
     \noalign
     {\kern1.5pt\nointerlineskip}
	 \ . \crcr\noalign{\kern-3pt
   }}}}\limits}

\def\theequation{\arabic{equation}}
\newcommand{\wh}{\widehat}
\newcommand{\wt}{\widetilde}
\newcommand{\wb}{\overline}
\newcommand{\ws}{\dot}
\newcommand{\nd}{{\phantom{\delta}}}
\newcommand{\sIm}[1]{\stackrel{#1}{\sim}}
\renewcommand{\th}[1]{\wh{\wt{#1}}}
\renewcommand{\S}{\phantom{\Big{|}}}
\newcommand{\aob}[2]{\frac{\textstyle{#1}}{\textstyle{#2}}}
\newcommand{\ssQ}{\mathcal{Q}}

\begin{document}
\paper{B\"acklund transformations for integrable lattice equations}
\author{James Atkinson}
\address{Department of Applied Mathematics, University of Leeds, Leeds LS2 9JT, UK}
\ead{james@maths.leeds.ac.uk}
\date{\today}
\begin{abstract}
We give new B\"acklund transformations (BTs) for some known integrable (in the sense of being multidimensionally consistent) quadrilateral lattice equations.
As opposed to the natural auto-BT inherent in every such equation, these BTs are of two other kinds.
Specifically, it is found that some equations admit additional auto-BTs (with B\"acklund parameter), whilst some pairs of apparently distinct equations admit a BT which connects them.
\end{abstract}

\section{Introduction}
Multidimensional consistency \cite{nw,bs} is the essence of integrability found in examples of lattice equations which arise as the superposition principle for B\"acklund transformations (BTs).
This property is deep enough to capture fully the integrability of a system, but manageable enough to be successfully employed in attempts to construct and classify integrable lattice equations \cite{abs1,abs2,hie1,hie2}.

In the present article, relationships between known examples of multidimensionally consistent equations are established.
These relationships are similar in spirit to the notion of multidimensional consistency.
However, rather than an equation being consistent with copies of itself, distinct equations are consistent with each other.
This consistency is equivalent to the existence of a particular kind of BT, and it is this latter point of view we adopt because it lends more in the way of intuition to the systems discussed.

The sense in which we use the term BT throughout this article, is for an overdetermined system in two variables which constitutes a transformation between solutions of the two equations that emerge as the compatibility constraints.
The term auto-BT will be used to describe the case where the emerging equations coincide.
We refer to a free parameter of an auto-BT as a B\"acklund parameter if transformations with different values of the parameter commute (in the sense that a superposition principle exists - examples will be given).

\section{The degenerate cases of Adler's equation}
A scalar multidimensionally consistent lattice equation of particular significance was found by Adler \cite{adler} as the superposition principle for BTs of the Krichever-Novikov equation \cite{KN1,KN2}.
We write Adler's equation in the following way,
\begin{equation}
p(u\wt{u}+\wh{u}\th{u})-q(u\wh{u}+\wt{u}\th{u}) = \frac{Qp-Pq}{1-p^2q^2}\left(u\th{u}+\wt{u}\wh{u}-pq(1+u\wt{u}\wh{u}\wt{\wh{u}})\right).
\label{q4}
\end{equation}
Here $u=u(n,m)$, $\wt{u}=u(n+1,m)$, $\wh{u}=u(n,m+1)$ and $\wh{\wt{u}}=u(n+1,m+1)$ denote values of the dependent variable $u$ as a function of the independent variables $n,m\in\mathbb{Z}$.
The {\it lattice parameters} $(p,P)$ and $(q,Q)$ are points on an elliptic curve, $(p,P),(q,Q)\in\Gamma$, 
\[
\Gamma = \left\{(x,X):X^2=x^4+1-(k+1/k) \ x^2\right\}
\label{eq:Gamma}
\]
where $k$ is an arbitrary constant (the Jacobi elliptic modulus).
The lattice parameters can be viewed as having their origin in B\"acklund parameters associated with commuting BTs of the Krichever-Novikov equation, they play a central role in integrability of (\ref{q4}).
Equation (\ref{q4}), the {\it Jacobi form} of Adler's equation, was first given by Hietarinta \cite{hie2}, it is equivalent (by a change of variables) to the {\it Weierstrass form} given originally by Adler \cite{adler}, cf. \cite{ahn1}.

Adler's equation was included in the list of multidimensionally consistent equations given later by Adler, Bobenko and Suris (ABS) in \cite{abs1} (where it was denoted $Q4$). 
Here we reproduce the remaining equations in that list:
\vspace{10pt}
\begin{equation}
\fl
\begin{array}{rrcl}
Q3^\delta :\! &\!(p\!-\!\aob{1}{p})(u\wt{u}+\wh{u}\th{u})\!-\!(q\!-\!\aob{1}{q})(u\wh{u}+\wt{u}\th{u}) &=& (\aob{p}{q}\!-\!\aob{q}{p})(\wt{u}\wh{u}+u\th{u}\!+\!\frac{\delta^2}{4}(p\!-\!\aob{1}{p})(q\!-\!\aob{1}{q})),\!\\
Q2^\nd :\! &p(u\!-\!\wh{u})(\wt{u}\!-\!\th{u})-q(u\!-\!\wt{u})(\wh{u}\!-\!\th{u}) &=& \!pq(q\!-\!p)(u\!+\!\wt{u}\!+\!\wh{u}\!+\!\th{u}\!-\!p^2\!+\!pq\!-\!q^2),\!\\
Q1^\delta:\! & p(u\!-\!\wh{u})(\wt{u}\!-\!\th{u})-q(u\!-\!\wt{u})(\wh{u}\!-\!\th{u}) &=& \delta^2 pq(q-p),\\
A2^\nd :\! &\!(p\!-\!\aob{1}{p})(u\wh{u}+\wt{u}\th{u})\!-\!(q\!-\!\aob{1}{q})(u\wt{u}+\wh{u}\th{u}) &=& (\aob{p}{q}-\aob{q}{p})(1+u\wt{u}\wh{u}\th{u}), \\
A1^\delta:\! & p(u\!+\!\wh{u})(\wt{u}\!+\!\th{u})-q(u\!+\!\wt{u})(\wh{u}\!+\!\th{u}) &=& \delta^2 pq(p-q),\\
H3^\delta:\! & p(u\wt{u}+\wh{u}\th{u}) - q(u\wh{u}+\wt{u}\th{u}) &=& \delta(q^2-p^2), \\
H2^\nd :\! &(u-\th{u})(\wt{u}-\wh{u})&=&(p-q)(u\!+\!\wt{u}\!+\!\wh{u}\!+\!\th{u}\!+\!p\!+\!q), \\
H1^\nd :\! &(u-\th{u})(\wt{u}-\wh{u})&=&(p-q), \\
\end{array}
\label{list}
\end{equation}
\vspace{10pt}
where it appears, $\delta$ is a constant parameter of the equation.
The equations in the list (\ref{list}) are all degenerate sub cases of the equation (\ref{q4}).
Table \ref{degen} contains the details of these degenerations.
\begin{table}[t]
\begin{center}
\begin{tabular}{c|c|c|c|c}
Eq$^{\nd}$ & $u$ & $k$ & $p$ &$P$ \\
\hline
$Q3^\delta\S$ & $\frac{2i\epsilon}{\delta}u$ & $-4\epsilon^2$ & $\epsilon(p-\frac{1}{p})$& $\frac{1}{2}(p+\frac{1}{p})+O(\epsilon^4)$\\
$Q2^\nd\S$ & $\frac{1}{\epsilon}+\frac{\epsilon}{2}u$ & $\epsilon^2$ & $\epsilon^2 p$& $1-\frac{\epsilon^2}{2}p^2-\frac{\epsilon^4}{8}p^4+O(\epsilon^6)$\\
$Q1^\delta\S$ & $\frac{\epsilon}{\delta}u$ & $k$ & $\epsilon p$& $1+O(\epsilon^2)$\\
$A2^\nd\S$ & $u$ & $-4\epsilon^2$ & $\frac{1}{\epsilon}(p-\frac{1}{p})^{-1}$ & $\frac{-1}{2\epsilon^2}(p+\frac{1}{p})(p-\frac{1}{p})^{-2}+O(\epsilon^2)$\\
$A1^\delta\S$ & $\frac{\epsilon}{\delta} u$ & $k$ & $\epsilon p$ & $-1+O(\epsilon^2)$\\
$H3^\delta\S$ & $1+\frac{\epsilon}{\sqrt{-\delta}}u$ & $1$ & $1-\frac{\epsilon^2}{2}p$& $-\epsilon^2p+O(\epsilon^4)$\\
$H2^\nd\S$ & $\frac{1}{\epsilon}+\epsilon-\frac{\epsilon}{2}u$ & $-4\epsilon^4$ & $1-\frac{\epsilon^2}{2}p$& $\frac{-1}{2\epsilon^2}+\frac{1}{4}p-2\epsilon^2+\epsilon^4p-\epsilon^6p+O(\epsilon^{10})$\\
$H1^\nd\S$ & $1+\epsilon u$ & $k$ & $1-\frac{\epsilon^2}{2}p$& $\frac{k-1}{\sqrt{-k}}-\epsilon^2\frac{k-1}{2\sqrt{-k}}p+O(\epsilon^4)$
\end{tabular}
\caption{{\it Substitutions which lead to the indicated degenerate sub case (Eq) of Adler's equation {\textrm (\ref{q4})} in the limit $\epsilon\rightarrow 0$. Choose $\delta=\epsilon$ rather than $0$ to arrive at Eq with $\delta=0$.}} \label{degen}
\end{center}
\end{table}
To clarify the meaning of the entries in this table we include an example here.
Let us make the substitutions
\[ u \rightarrow \epsilon u, \quad p\rightarrow \epsilon p, \quad q\rightarrow\epsilon q \]
in (\ref{q4}) and consider the leading term in the small-$\epsilon$ expansion of the resulting expression.
For this calculation it is necessary to write the parameters $P$ and $Q$ as a series in $\epsilon$, 
\[\fl P = \pm(1-\epsilon^2\frac{1}{2}(k+1/k)p^2+\ldots), \quad Q = \pm(1-\epsilon^2\frac{1}{2}(k+1/k)q^2+\ldots), \]
so there is some choice of sign.
The rest of the calculation is straightforward and the leading order expression that results is exactly the equation $Q1^1$ or $A1^1$ depending on this choice of sign.

It was pointed out in \cite{abs1} that one can descend through the lists `Q', `A' and `H' in (\ref{list}) by degeneration from $Q4$, $A2$ and $H3^\delta$ respectively.
The degenerations from Adler's equation in Weierstrass form to the equations in the `Q' list are given explicitly in \cite{q4}.

Part of what gives (\ref{q4}) its particular significance is that, as far as we are aware, all known scalar multidimensionally consistent lattice equations are either linearisable or transformable to (\ref{q4}) or one of its degenerate sub cases (\ref{list}) (possibly by a non-autonomous, or gauge, transformation).
Note, this apparent ubiquity of Adler's equation is partially explained by the main result in \cite{abs2}.

\section{Alternative auto-B\"acklund transformations}
Table \ref{autoBTlist} lists auto-BTs for some particular equations from the list (\ref{list}).
The BTs listed are distinct from the natural auto-BT associated with every multidimensionally consistent equation (for example, this is described for Adler's equation in \cite{ahn1}), one significant difference is that the superposition principle associated with these alternative auto-BTs coincides with some other equation present in the list (\ref{list}).

To explain the implementation of the BTs in table \ref{autoBTlist} we give an example here (the last entry in the table).
\begin{table}[t]
\begin{center}
\begin{tabular}{c|c|c}
Eq & \S B\"acklund transformation \S & SP\\
\hline
$Q3^0$ & \S $(pr-\aob{1}{pr})(uv+\wt{u}\wt{v})-(r-\aob{1}{r})(u\wt{v}+\wt{u}v)=(p-\aob{1}{p})(1+u\wt{u}v\wt{v})$ & $A2^\nd$\\
$Q1^\delta$ & \S $p(u+v)(\wt{u}+\wt{v})-r(u-\wt{u})(v-\wt{v})=\delta^2pr(p+r)$ & $A1^\delta$\\
$Q3^0$ & \S $p(u\wt{v}+\wt{u}v)-uv-\wt{u}\wt{v}=\delta r(1-p^2)$ & $H3^\delta$\\
$Q1^1$ & \S $(u-\wt{u})(v-\wt{v})=-p(u\!+\!\wt{u}\!+\!v\!+\!\wt{v}\!+\!p\!+\!2r)$ & $H2^\nd$\\
\end{tabular}
\caption{{\it Each equation, Eq, admits the given auto-BT (with B\"acklund parameter $r$). The equation SP emerges as the superposition principle for solutions of Eq related by this BT. It turns out that the converse associations also hold (see main text).} \label{autoBTlist}}
\end{center}
\end{table}
Consider the following system of equations in the two variables $u(n,m)$ and $v(n,m)$,
\begin{equation}
\begin{array}{rcl}
(u-\wt{u})(v-\wt{v}) & = & -p(u+\wt{u}+v+\wt{v}+p+2r),\\
(u-\wh{u})(v-\wh{v}) & = & -q(u+\wh{u}+v+\wh{v}+q+2r)
\end{array}
\label{exaBT}
\end{equation}
(the second equation here is implicit from the first and so is omitted from the table for brevity).
With $u$ fixed throughout the lattice (i.e., for all $n,m$), (\ref{exaBT}) constitutes an overdetermined system for $v$.
This is resolved ($\wt{\wh{v}}=\wh{\wt{v}}$) if $u$ is chosen to satisfy the equation $Q1^1$ throughout the lattice, moreover, $v$ which then emerges in the solution of (\ref{exaBT}) also satisfies $Q1^1$.
We say that the solutions $u$ and $v$ of $Q1^1$ are related by the BT (\ref{exaBT}) and for convenience write 
\begin{equation}
u\sIm{r} v.
\label{btrel}
\end{equation}
Here $r$ is the parameter present in (\ref{exaBT}), this is a free parameter of the transformation.
The relation (\ref{btrel}) is symmetric because (\ref{exaBT}) is invariant under the interchange $u\leftrightarrow v$.

Transformations (\ref{exaBT}) with different choices of the parameter $r$ commute in the sense that a superposition principle exists.
That is, given a solution $u(n,m)$ of $Q1^1$, suppose we compute other solutions $\wb{u}(n,m)$, $\ws{u}(n,m)$, $\ws{\wb{u}}(n,m)$ and $\wb{\ws{u}}(n,m)$ for which
\begin{equation}
\begin{array}{c}
u \sIm{r} \wb{u}, \qquad \wb{u} \sIm{s} \ws{\wb{u}},\\
u \sIm{s} \ws{u}, \qquad \ws{u} \sIm{r} \wb{\ws{u}}.
\end{array}
\label{btrels}
\end{equation}
Then the solutions $\ws{\wb{u}}$ and $\wb{\ws{u}}$ coincide throughout the lattice provided they coincide at a single point where the equation
\begin{equation}
(u-\ws{\wb{u}})(\wb{u}-\ws{u}) = (r-s)(u+\wb{u}+\ws{u}+\ws{\wb{u}}+r+s)
\label{exspf}
\end{equation}
also holds (and in the computation of these new solutions we can always choose the integration constants to make this so).
Furthermore, the relation (\ref{exspf}) then continues to hold throughout the lattice.
In this sense we regard (\ref{exspf}) as the superposition principle for solutions of $Q1^1$ related by the BT (\ref{exaBT}), up to a change in notation (\ref{exspf}) coincides with the lattice equation $H2$ from the list (\ref{list}).

To conclude our description of the BT (\ref{exaBT}) we recognise that the preceding facts are also true in the converse sense.
Observe first that the system (\ref{btrels}) implies (amongst others) the following equations,
\begin{equation}
\begin{array}{rcl}
(u-\wt{u})(\wb{u}-\wb{\wt{u}}) & = & -p(u+\wt{u}+\wb{u}+\wb{\wt{u}}+p+2r),\\
(u-\wt{u})(\ws{u}-\ws{\wt{u}}) & = & -p(u+\wt{u}+\ws{u}+\ws{\wt{u}}+p+2s).
\end{array}
\label{exaBT2}
\end{equation}
Now consider (\ref{exspf}) as a lattice equation, so that $u=u(l,k)$, $\wb{u}=u(l+1,k)$, $\ws{u}=u(l,k+1)$ and $\ws{\wb{u}}=u(l+1,k+1)$ for new independent variables $l,k\in\mathbb{Z}$.
Then the system (\ref{exaBT2}) forms a BT, with B\"acklund parameter $p$, between solutions $u(l,k)$ and $\wt{u}(l,k)$ of the equation (\ref{exspf}) (i.e., constitutes an auto-BT for $H2$).
This BT commutes with its counterpart with B\"acklund parameter $q$ (which relates solutions $u(l,k)$ and $\wh{u}(l,k)$ of (\ref{exspf})), the superposition principle in this case is exactly the equation $Q1^1$. 

As described for this example, all the BTs given in table \ref{autoBTlist} establish a kind of duality between a particular pair of equations, specifically, one equation emerges as the superposition principle for BTs that relate solutions of the other.
This can be compared to the natural auto-BT of a multidimensionally consistent equation, for which the superposition principle coincides with the equation itself.

\section{B\"acklund transformations between distinct equations}
Table \ref{BTlist} lists BTs that connect particular pairs of equations from the list (\ref{list}).
\begin{table}[b]
\begin{center}
\begin{tabular}{c|c|c}
Eq in $u$ & \S B\"acklund transformation \S & Eq in $v$ \\
\hline
$Q3^0$ & \S $uv+\wt{u}\wt{v}-p(u\wt{v}+\wt{u}v)=(p-\aob{1}{p})(u\wt{u}+\frac{\delta^2\! }{4} p)$ & $Q3^\delta$\\
$Q1^1$ & \S $(u-\wt{u})(v-\wt{v})=p(v+\wt{v}-2u\wt{u})+p^2(u+\wt{u}+p)$ & $Q2\ $\\
$Q1^0$ & \S $(u-\wt{u})(v-\wt{v})=p(u\wt{u}-\delta^2)$ & $Q1^\delta$ \\
$H3^0$ & \S $pu\wt{u}-uv-\wt{u}\wt{v} = \delta$ & $H3^\delta$ \\
$H1\ $ & \S $2u\wt{u}=v+\wt{v}+p$ & $H2\ $\\
$A1^0$ & \S $(u+\wt{u})(v+\wt{v})=p(u\wt{u}+\delta^2)$ & $A1^\delta$ \\
$A1^0$ & \S $(u+\wt{u})(v-\wt{v})=p (u-\wt{u})$ & $Q1^1$ \\
$\!{}^\dagger A1^\delta$ & \S $u+\wt{u} = 2pv\wt{v} + \delta p^2$ & $H3^\delta$ \\
$\!{}^\dagger A1^0$ & \S $(u+\wt{u})v\wt{v}=p(1-\frac{\delta}{2}u)(1-\frac{\delta}{2}\wt{u})$ & $H3^\delta$ \\
\end{tabular}
\caption{{\it 
BTs between distinct lattice equations. 
The BT between $Q1^0$ and $Q1^\delta$ was given originally by ABS in \cite{abs2}.
$\dagger$ indicates application of the point transformation $p\rightarrow p^2$, $q\rightarrow q^2$ to the lattice parameters. 
\label{BTlist}}}
\end{center}
\end{table}
To be precise about the meaning of the entries in this table we again give an example.
Consider the system of equations
\begin{equation}
\begin{array}{rcl}
2u\wt{u}=v+\wt{v}+p,\\
2u\wh{u}=v+\wh{v}+q.
\end{array}
\label{exBT}
\end{equation}
This system is compatible in $v$ if the variable $u$ satisfies the equation $H1$.
Given such $u$, it can be verified that $v$ which emerges in the solution of (\ref{exBT}) then satisfies the equation $H2$.
Conversely, if $v$ satisfies $H2$ then solving (\ref{exBT}) yields $u$ which satisfies $H1$.
In this way the system (\ref{exBT}) constitutes a BT between the equations $H1$ and $H2$, which corresponds to the fifth entry in table \ref{BTlist} (where we give only one equation from the pair (\ref{exBT}), the other being implicit).

The BT (\ref{exBT}) can be explained as a non-symmetric degeneration of the natural auto-BT for the equation $H2$, which is defined by the system
\begin{equation}
\begin{array}{rcl}
(u-\wt{v})(\wt{u}-v) &=& (p-r)(u+\wt{u}+v+\wt{v}+p+r),\\
(u-\wh{v})(\wh{u}-v) &=& (q-r)(u+\wh{u}+v+\wh{v}+q+r) 
\end{array}
\label{H2BT}
\end{equation}
($r$ is the B\"acklund parameter).
Now, the substitution $u\rightarrow\frac{1}{\epsilon^2}+\frac{2}{\epsilon}u$ in the equation $H2$ leads to the equation $H1$ in the limit $\epsilon\longrightarrow 0$.
This substitution in the system (\ref{H2BT}) together with the particular choice $r=-\frac{1}{\epsilon^2}$ yields the system (\ref{exBT}) in the limit $\epsilon\longrightarrow 0$.
Note that it is not a priori obvious that the BT will be preserved in this limit, by which we mean that once the system (\ref{exBT}) has been found, it remains to verify the result.

It can be confirmed that all but the last three entries in table \ref{BTlist} are explained as a non-symmetric degeneration of the natural auto-BT for the equation in $v$, however the last three entries have been found by ad hoc methods.
The transformations in table \ref{BTlist} are stated up to composition with point symmetries of the equations in $u$ and $v$.

\vspace{12pt}

To conclude this section we give two more BTs.
These connect multidimensionally consistent lattice equations which lie outside the list (\ref{list}).
Consider first the system (with 2-component lattice parameters)
\begin{equation}
\begin{array}{rcl}
(u+p_1)v &=& (\wt{u}+p_2)\wt{v},\\
(u+q_1)v &=& (\wh{u}+q_2)\wh{v}.
\end{array}
\label{hbt}
\end{equation}
This constitutes a BT between the pair of lattice equations
\begin{eqnarray}
(u\!+\!q_1)(\wt{u}\!+\!p_2)(\wh{u}\!+\!p_1)(\th{u}\!+\!q_2) = (u\!+\!p_1)(\wh{u}\!+\!q_2)(\wt{u}\!+\!q_1)(\th{u}\!+\!p_2),
\label{hie}\\
(p_1\!-\!q_1)v + (p_2\!-\!q_2)\th{v} = (p_2\!-\!q_1)\wt{v} + (p_1\!-\!q_2)\wh{v}.
\label{lin}
\end{eqnarray}
The equation (\ref{hie}) was given originally by Hietarinta in \cite{hie1} and subsequently shown to be linearisable by Ramani et al. in \cite{dhe}.
The BT (\ref{hbt}) provides an alternative linearisation by connecting it with the equation (\ref{lin}).

The other example is a BT between equations of rank-2 (i.e., 2 component systems) and is therefore outside the list given by ABS \cite{abs1} where only scalar equations are considered.
It is defined by the system (with scalar lattice parameters)
\begin{equation}
\begin{array}{cc}
(v_1-\wt{v}_1)u_2\wt{u}_2 = pu_1, & \qquad (v_2-\wt{v}_2)u_1\wt{u}_1=p\wt{u}_2,\\
(v_1-\wh{v}_1)u_2\wh{u}_2 = qu_1, & \qquad (v_2-\wh{v}_2)u_1\wh{u}_1=q\wt{u}_2,
\end{array}
\label{bsqBT}
\end{equation}
and connects the equations
\begin{equation}
\begin{array}{c}
p(u_2\wt{u}_2\wh{u}_1-u_1\wh{u}_2\th{u}_2) = q(u_2\wh{u}_2\wt{u}_1-u_0\wt{u}_2\th{u}_2),\\
p(u_1\wt{u}_1\th{u}_2-\wt{u}_2\wh{u}_1\th{u}_1) = q(u_1\wh{u}_1\th{u}_2-\wh{u}_2\wt{u}_1\th{u}_1),
\end{array}
\label{mbsq}
\end{equation}
and
\begin{equation}
\begin{array}{c}
p^3(v_1-\wh{v}_1)(\wt{v}_1-\th{v}_1)(\wt{v}_2-\th{v}_2) = q^3(v_1-\wt{v}_1)(\wh{v}_1-\th{v}_1)(\wh{v}_2-\th{v}_2),\\
p^3(v_2-\wh{v}_2)(\wt{v}_2-\th{v}_2)(v_1-\wh{v}_1) = q^3(v_2-\wt{v}_2)(\wh{v}_2-\th{v}_2)(v_1-\wt{v}_1).
\end{array}
\label{sbsq}
\end{equation}
The equation (\ref{mbsq}) is the lattice modified Boussinesq equation given originally as a second order scalar equation in \cite{npcq}, the rank-2 version (\ref{mbsq}) is attributable to Nijhoff in \cite{nij1}.
The equation (\ref{sbsq}) is a rank-2 version of the lattice Schwarzian Boussinesq equation which was given originally as a second order scalar equation in \cite{nij2} (a second-order scalar equation can be recovered from (\ref{mbsq}) or (\ref{sbsq}) by elimination of one of the variables from the two-component system, by second order here we mean a lattice equation on a square nine point stencil).

The BT (\ref{bsqBT}) naturally generalises a scalar BT given in \cite{nrgo} which connects the lattice modified and Schwarzian Korteweg-de Vries equations.
(Note, when transformed to a BT between equations from the list (\ref{list}) this becomes a non-autonomous BT, a type of BT not considered in the present article.)

\section{Discussion}
In the preceding sections we have given systems of equations which may be written generically in the form
\begin{equation}
\begin{array}{rcl}
f_p(u,\wt{u},v,\wt{v}) &=& 0,\\
f_q(u,\wh{u},v,\wh{v}) &=& 0,
\end{array}
\label{gBT}
\end{equation}
and that constitute a BT between a lattice equation in $u=u(n,m)$ and a possibly different lattice equation in $v=v(n,m)$, say
\begin{eqnarray}
\ssQ_{pq}(u,\wt{u},\wh{u},\th{u}) &=& 0,
\label{geq1}\\
\ssQ^*_{pq}(v,\wt{v},\wh{v},\th{v}) &=& 0.
\label{geq2}
\end{eqnarray}
(Here we suppose that $u$ and $v$ are scalar fields and $f$, $\ssQ$ and $\ssQ^*$ are polynomials of degree 1 in which the coefficients are functions of the lattice parameters.)
In this generic (scalar) case it can be deduced (by considering an initial value problem on the cube) that the multidimensional consistency of (\ref{geq1}) implies the multidimensional consistency of (\ref{geq2}).
Furthermore, when (\ref{geq1}) and (\ref{geq2}) are multidimensionally consistent, the BT (\ref{gBT}) commutes with the natural auto-BTs for these equations, the superposition principle being the equation
\begin{equation}
f_r(u,\wb{u},v,\wb{v}) \ = \ 0.
\label{gspf}
\end{equation}
Here $u$ and $\wb{u}$ are solutions of (\ref{geq1}) related by its natural auto-BT (with B\"acklund parameter $r$), similarly $v$ and $\wb{v}$ are solutions of (\ref{geq2}) related by its natural auto-BT (also with B\"acklund parameter $r$), and finally, $u$ and $\wb{u}$ are related to $v$ and $\wb{v}$ respectively by the BT (\ref{gBT}).

We remark that not all lattice equations (\ref{geq1}), (\ref{geq2}) which arise in this way are multidimensionally consistent. 
Consider the following example (which involves 2-component lattice parameters),
\begin{equation}
\begin{array}{rcl}
p_1u\wt{u} &=& v+\wt{v}+p_2,\\
q_1u\wt{u} &=& v+\wt{v}+q_2.
\end{array}
\label{ncacBT}
\end{equation}
This system constitutes a BT between the equations
\begin{equation}
\begin{array}{l}
p_1(u\wt{u}+\wh{u}\th{u})-q_1(u\wh{u}+\wt{u}\th{u}) = 2(p_2-q_2),\\
p_1^2(v+\wh{v})(\wt{v}+\th{v})-q_1^2(v+\wt{v})(\wh{v}+\th{v}) = p_2^2q_1^2-q_2^2p_1^2.
\end{array}
\label{ncac}
\end{equation}
The equations (\ref{ncac}) are multidimensionally consistent if and only if the components of the lattice parameters are connected by the relations
\begin{equation} a+bp_1^2+cp_2 = 0, \qquad a+bq_1^2+cq_2=0, \label{cond} \end{equation}
for some constants $a,b$ and $c$ not all equal to zero. 
(The solution of (\ref{cond}) yields the fifth and eighth entries in table \ref{BTlist}.)
On the other hand, when (\ref{gBT}) constitutes an auto-BT, so that $\ssQ^*=\ssQ$, we have found no counterexamples to the conjecture that the equation defined by $\ssQ$ is multidimensionally consistent.

\section{Concluding remarks}
The B\"acklund transformations (BTs) given in this article establish new relationships between equations within the classification of Adler, Bobenko and Suris (ABS) \cite{abs1}.
Alternative auto-BTs turn out to establish a kind of duality between some pairs of equations.
Transformations connecting other pairs of equations are of practical significance, for example allowing for soliton solutions to be found for one equation from those of the other (cf. \cite{ahn2}).

New BTs have also been established for integrable lattice equations which lie outside the classification of ABS. 
In particular we give a BT between systems of rank-2 where only a few examples of multidimensionally consistent equations are known.

\ack
The author was supported by the UK Engineering and Physical Sciences Research Council (EPSRC) and is indebted to Frank Nijhoff for his continued guidance.

\end{document}